# Simulating Radiation Shielding Effectiveness Against Three Neutron Sources

Andrew K. Gillespie [a]*, Cuikun Lin [a], Matthew Looney[a*], and R. V. Duncan [a]

## AFFILIATIONS

[a]*Department of Physics and Astronomy, Texas Tech University, Lubbock Texas 79409, USA*
*Corresponding authors: a.gillespie@ttu.edu and matthew.looney@ttu.edu
MCNP simulations were conducted by Dr Andrew Gillespie, licensed by RSICC.

**Keywords: neutron, radiation, shielding, boric acid, concrete, light water, heavy water, californium**

## ABSTRACT

Laboratories and Universities regularly apply for approval from the United States Nuclear Regulatory Commission (NRC) to use neutron generators for experimental research. To comply with the regulations set by the NRC, adequate shielding is necessary to ensure that the radiation dose rates experienced by an operator, and outside the walls of any containment buildings, are below the prescribed levels. Typically, the neutron source needs to be shielded such that the radiation dose rate experienced by any user is less than 0.25 mRem/hr (500mrem/2000-hour work-year)[1]. To address this requirement, we investigate the effectiveness of boronated concrete, boronated water, and light water shielding materials and their applicability to three neutron sources. We present our findings on the radiation shielding design and calculations for three neutron sources situated inside shielding layers. Our modeling utilized the Monte Carlo n-Particle transport codes (MCNP6.2)[2,3] to simulate neutron attenuation of the shielding. The simulation results reveal that a light water shielding can sufficiently reduce the dose rate for an individual located as close as two meters from the source. Therefore, this shield design can effectively decrease the radiation dose below the maximum recommended limit.

## I. SIMULATION INTRODUCTION

Experiments involving nuclear fission or fusion require strict adherence to safety protocols. As a first step, the radiation dose rate needs to be simulated for specific sets of experiments. To comply with the regulations set by the United States NRC, adequate shielding is necessary to ensure that the radiation dose rates experienced are below the prescribed levels. Typically, the neutron source needs to be shielded such that the radiation dose rate outside of the shielding is less than 0.25 mRem/hr (500mRem/2000-hour work-year).[1] To address this requirement, we investigate the effectiveness of boronated concrete, boronated water, and light water shielding materials and their applicability to three neutron sources. Specifically, we investigate the P385 neutron generator model that generates 14.1 MeV neutrons isotropically at an approximate rate of $3x10^8$ n/s,[4] the DD-109 model that generates 2.45 MeV neutrons isotropically at an approximate rate of $10^9$ n/s,[5] and a 10-mCi Californium source with a peak neutron intensity around 2 MeV and approximate neutron generation of $4.4x10^7$ n/s. These simulations are applicable to any experiment involving a neutron generator in a laboratory environment.

Initial simulations of the equivalent dose rate were performed for operators near a neutron source inside a 30'x60' warehouse. Dose rates were simulated using MCNP6.2 for three locations within the warehouse.

MCNP has a plotter available that is capable of displaying cross sections of the geometry along primary axes. **Figure 1** displays three cross sections of the warehouse room with the neutron source contained in its center. The warehouse is 30 feet deep and 60 feet across. This simplified transport simulation involves a concrete floor and uses two sheets of drywall used for the outer walls. A hallow container of polyethylene is used to represent the walls of the shielding. The containers are filled with the 4% boric acid solution in light water. Three points are marked in the figure that represent the tally locations used to score the effective dose rate. P1 and P2 indicate the locations people might be standing if respectively located immediately outside of the 30-foot and 60-foot-wide length of the warehouse. P3 indicates the location of a potential operator who is standing 2 meters away from the source location. These tally regions are rectangular boxes filled with water and are 1’x1’x6’. The isotropic source is located at the center of the shielding box and is 90 cm above the cement floor. Though we will use six separate shielding walls, they are represented here as one connected volume for simplicity.

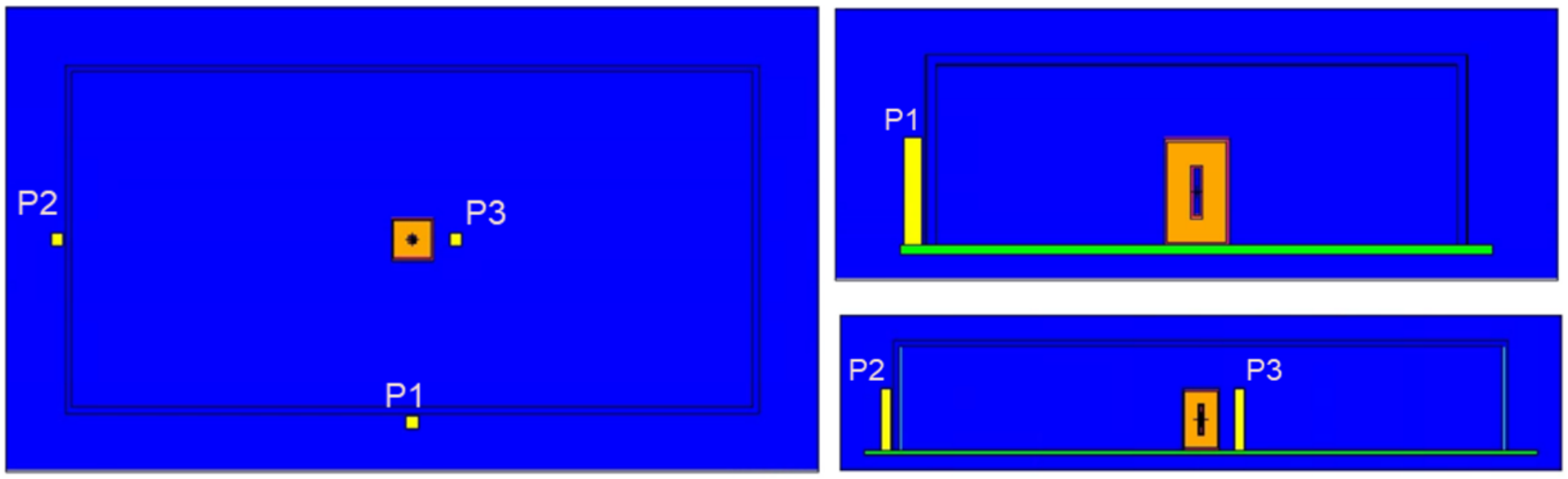

**Figure 1:** Cross-sections of the warehouse in which the neutron generator will be housed. *Left*: X-Y horizontal plane at a height of 90 cm. *Top Right*: Y-Z cross-section through the middle along the 30-foot-deep section of the warehouse. *Bottom Right*: X-Z cross-section through the middle of the 60-foot-wide section of the warehouse.

The shielding capability of the material layers was simulated for various thicknesses. The average track length estimate of the particle flux was tallied in the three regions of interest. Multiplying this by the source strength and dividing by the equivalent dose yields the effective dose in units of mRem/Hr. This helps to determine the minimum thickness of each material that would be required to bring the equivalent dose below acceptable limits.

As applied to a staffed laboratory setting, the equivalent dose experienced by the workers during regular operation needs to be minimized. The neutron source will be closest to the workers when maintenance is being performed immediately outside of the shielding, hence the equivalent dose may be approximated by the dose experienced at the location of P3 in the simulations in preparation for these laboratory experiments.

The average track length estimate of the particle flux was tallied in the three regions on interest. Multiplying this by the source strength and dividing it by the equivalent dose yields the effective dose in units of mRem/Hr. The most relevant tally zone is P3 since it represents an operator within a laboratory setting.

One known obstacle in using the average track length estimate of the particle flux is that large uncertainties can arise for tally regions that do not subtend a significant fraction of the solid angle from an isotropic source.[2] For an isotropic source, very few source particles will reach tally zones that are far away from the source. To reduce these uncertainties, we also simulated an equivalent tally zone that is the same distance from the isotropic source, but surrounds it as a cylindrical shell. In the simplified model, the shielding material is constructed as a cylindrical shell of various thicknesses and the P3 tally zone consists of a 1-foot-thick cylindrical shell that surrounds the shielding and isotropic neutron point source. The simplified model increases the statistical certainty of the tally results, and therefore increases the certainty of the calculated equivalent dose. The equivalent dose rates obtained from the simplified model have been compared with the P3 tally zone.

## II. CALCULATING THE EQUIVALENT DOSE

Data used in calculating the dose rates included the Mean Quality Factors, Q, and Fluence Per Unit Dose Equivalent for Monoenergetic Neutrons. This data was taken from the US NRC and can also be found in the supplementary materials. The fluence per unit dose equivalent considers (a) Value of quality factor (Q) at the point where the dose equivalent is maximum in a 30-cm diameter cylinder tissue-equivalent phantom and (b) Monoenergetic neutrons incident normally on a 30-cm diameter cylinder tissue-equivalent phantom.[6,7]

To calculate the dose rate, the neutron flux and energies were established throughout the three locations P1, P2 and P3 using the MCNP6.2 mesh tally feature. **Figure 2** displays the neutron tallies in the three different regions indicated in **Figure 1** when different thicknesses of light water shielding are used. The average track length estimate of the particle flux is reduced by a factor of approximately 100 when using a shielding thickness of 80 cm or more compared to average track length of the particle flux when using a shielding of only 0.1 cm.

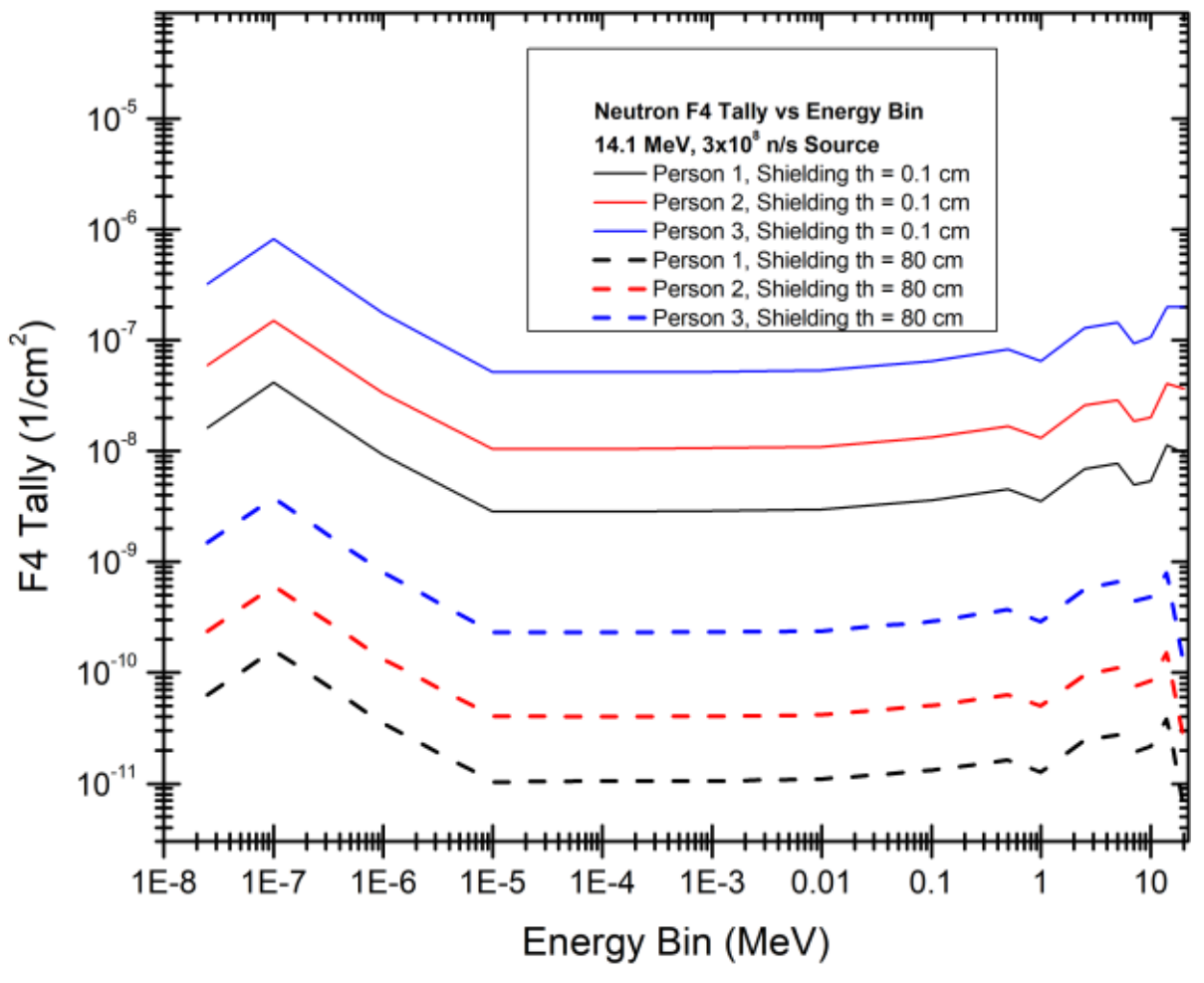

**Figure 2:** The average track length estimate of the particle flux at the locations of each tally regions.

**Figure 2** displays the neutron tallies in the three different regions indicated in **Figure 1** when different thicknesses of light water shielding are used. The average track length estimate of the particle flux is reduced by a factor of approximately 100 when using a shielding thickness of 80 cm or more compared to the shielding made with 0.1 cm of light water.

Additional calculations of equivalent dose rate were carried out, based on the NRC data presented in the supplementary materials and the mesh tally calculation shown in **Figure 2**. **Figure 3** illustrates the equivalent dose rates for Person 1, Person 2, and Person 3, respectively situated outside the 30 ft wall, outside the 60 ft wall, and 2 meters away from the source inside the warehouse, as a function of the shielding thickness.

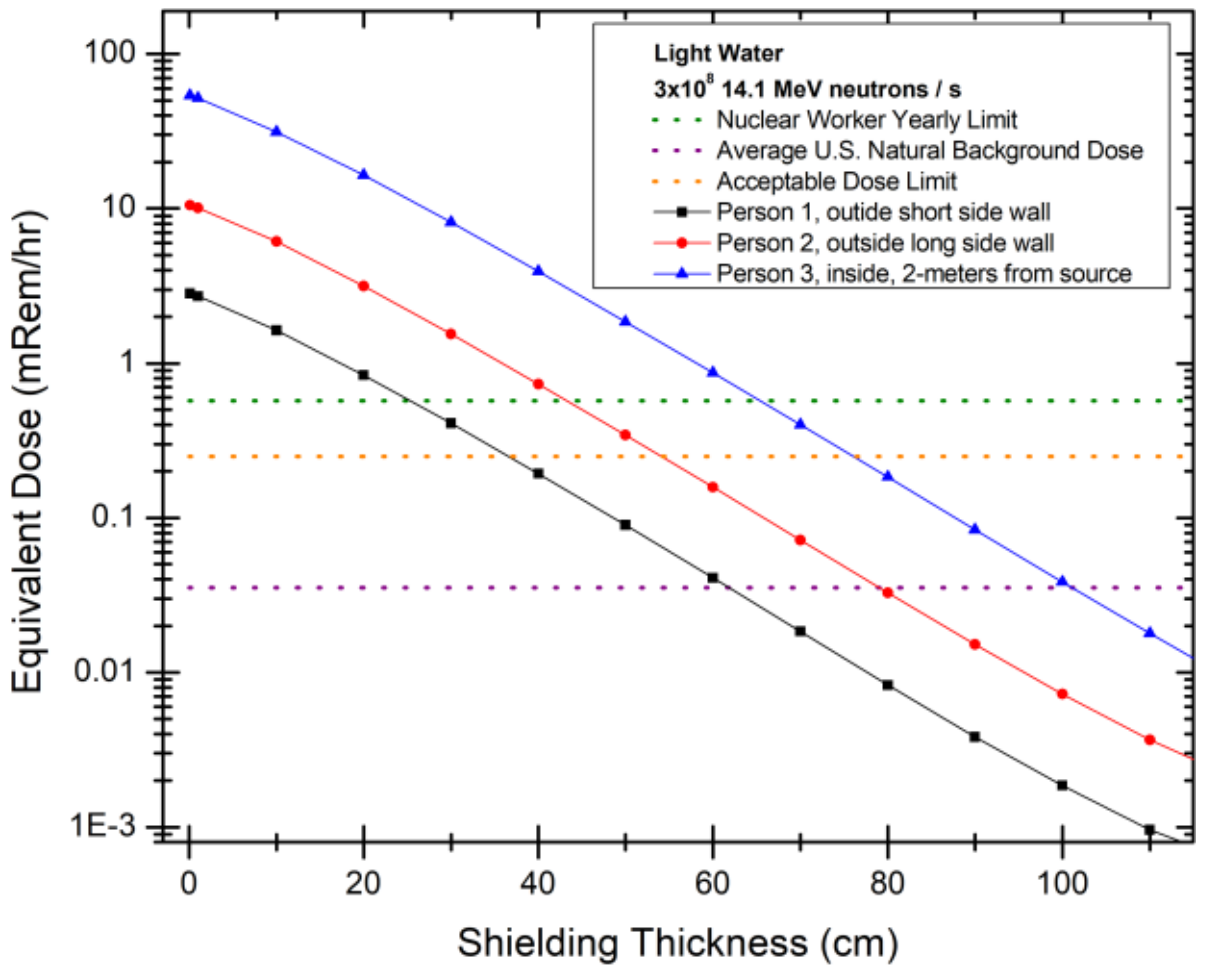

**Figure 3:** Equivalent dose rates for Person 1, Person 2, and Person 3 respectively located outside the 30 ft wall of the warehouse, outside the 60 ft wall of the warehouse, and inside the warehouse 2 meters away from the source as a function of shielding thickness of light water. The orange dotted line represents the acceptable limit of 0.25 mRem/h.

As an example, the results for light water indicate that less than 80 cm of shielding thickness is sufficient to reduce the source dose rate below the acceptable dose rate of 0.25 mRem/hr, demonstrating the effectiveness of our light water contained in polyethylene shielding design. As a result, the source's dose rate can be easily reduced to acceptable levels, summarized in **Table 1**. These series of simulations and calculations were performed for each neutron source and shielding design type within this article. We discuss the necessary minimum shielding thickness of these three materials that will be sufficient to reduce the equivalent radiation dose below the acceptable threshold.

## III. RESULTS AND DISCUSSION

### IIIa. Equivalent Dose Rate for a 14.1-MeV Neutron Source

**Figure 4** shows that the results for the P3 tally region and the simplified cylindrical model agree well for all results above the equivalent dose rate limit of 0.25 mRem/h. The orange dotted line represents the acceptable limit of 0.25 mRem/h. The tally results for the simplified cylindrical model pass all ten of the statistical checks within MCNP6.2, which indicates a high degree of validity of the simulation results.

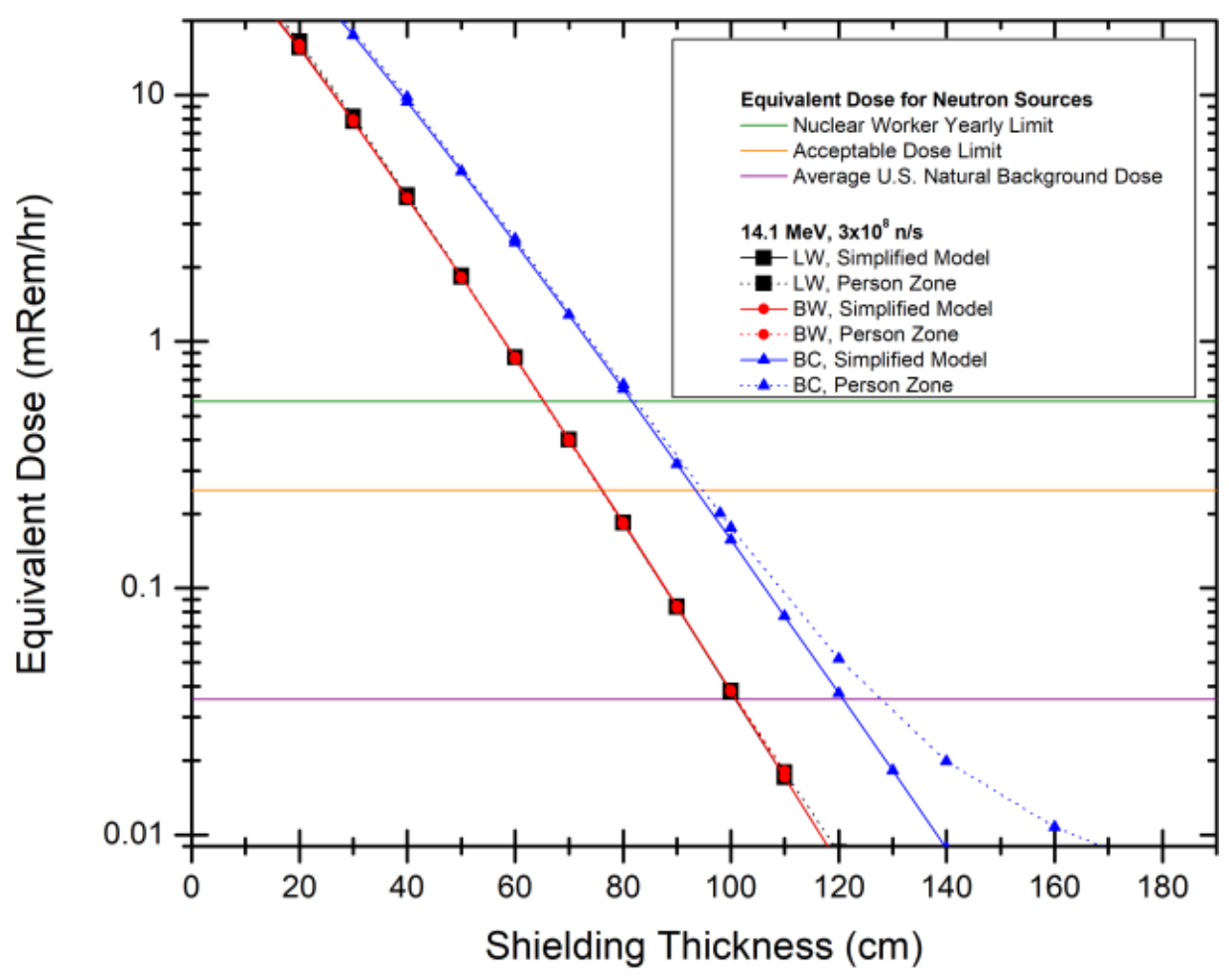

**Figure 4:** Calculated equivalent dose rate experienced by the P3 tally region when using various thicknesses of each shielding type for a 14.1 MeV neutron source.

Shielding types 1 & 3, containing either boric acid in water or containing only light water, resulted in similar shielding effectiveness. The dose rates are also in agreement between the P3 zone from the experimental warehouse and the simplified model. The equivalent dose rate limit is attained when approximately 77.1 cm of boric acid in water shielding is used, 94.3 cm of boric acid in concrete shielding is used, or 77.0 cm of light water shielding is used.

## IIIb. Equivalent Dose Rate for a 2.45 MeV Neutron Source

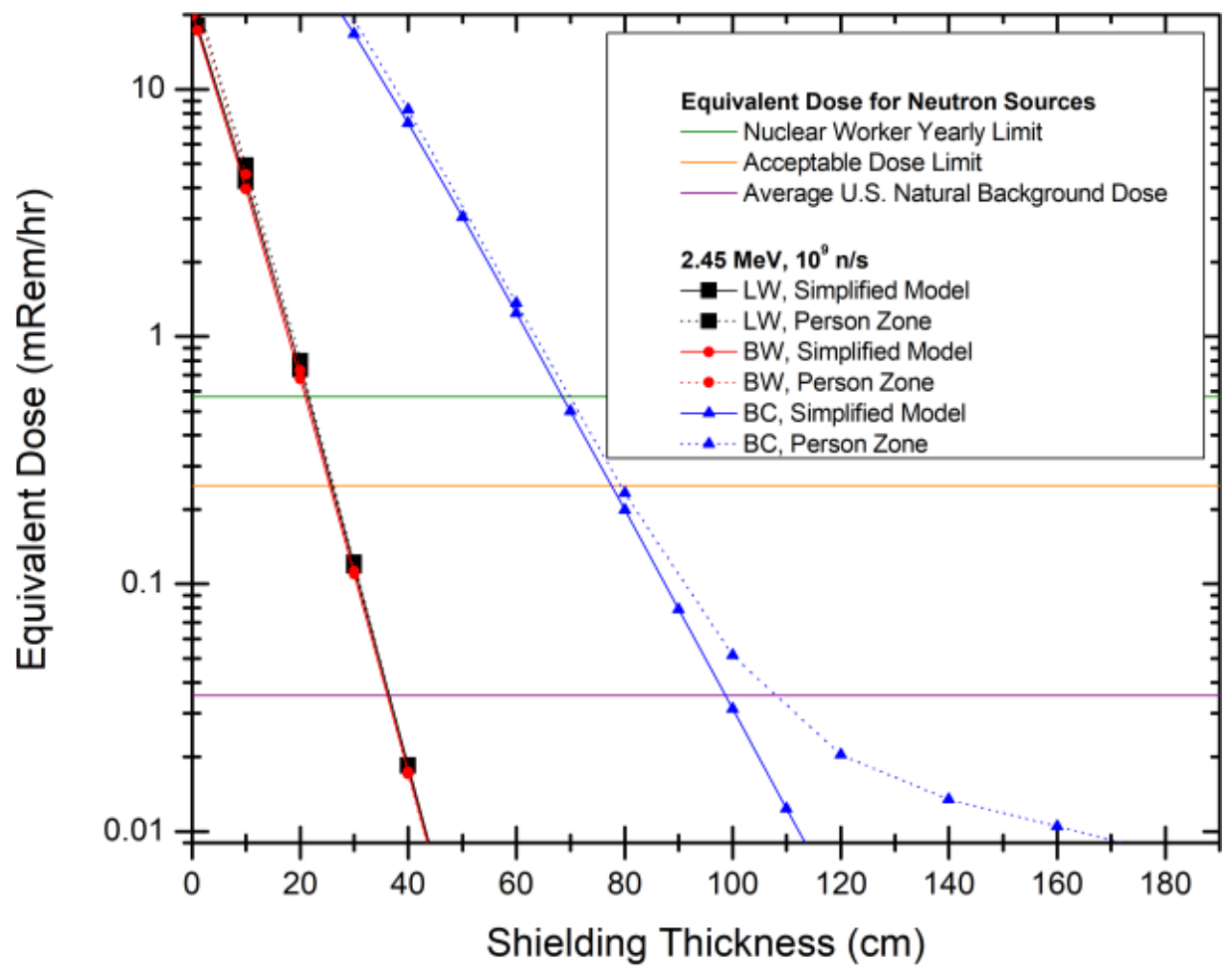

**Figure 5:** Calculated equivalent dose rate experienced by the P3 tally region when using various thicknesses of each shielding type for a 2.45 MeV neutron source.

**Figure 5** shows that the results for the P3 tally region and the simplified cylindrical model agree well for all results above the equivalent dose rate limit of 0.25 mRem/h. The orange dotted line represents the acceptable limit of 0.25 mRem/h. The tally results for the simplified cylindrical model pass all ten of the statistical checks within MCNP6.2, which indicates a high degree of validity of the simulation results. Shielding types 1 & 3, containing either boric acid in water or containing only light water, resulted in similar shielding effectiveness. The dose rates are also in agreement between the P3 zone from the experimental warehouse and the simplified model. The equivalent dose rate limit is attained when approximately 27.6 cm of boric acid in water shielding is used, 78.4 cm of boric acid in concrete shielding is used, or 27.9 cm of light water shielding is used.

### IIIc. Equivalent Dose Rate for a 10-mCi $^{252}Cf$ Source

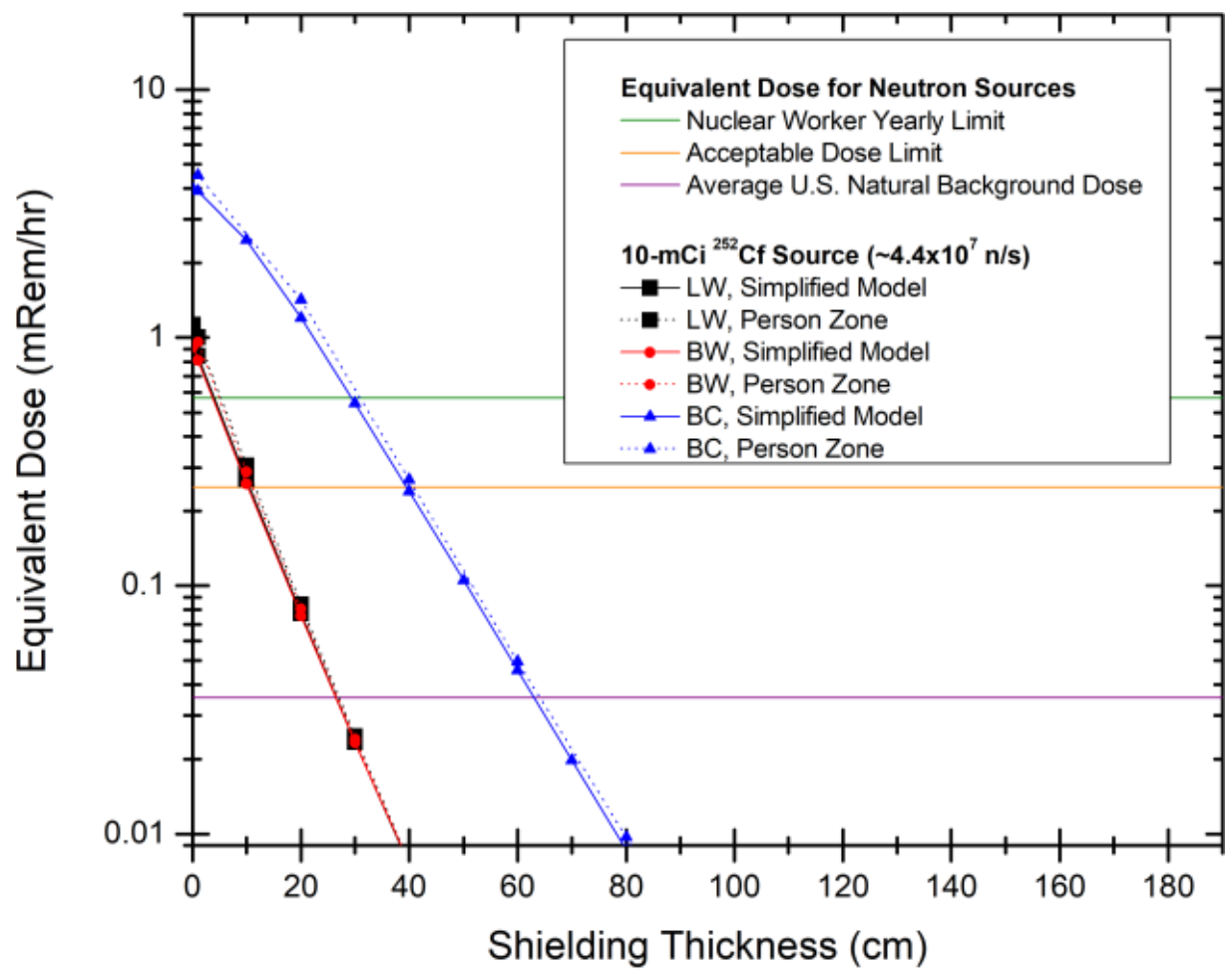

**Figure 6:** Calculated equivalent dose rate experienced by the P3 tally region when using various thicknesses of each shielding type for a 10-mCi $^{252}Cf$ neutron source.

**Figure 6** shows that the results for the P3 tally region and the simplified cylindrical model agree well for all results above the equivalent dose rate limit of 0.25 mRem/h. The orange dotted line represents the acceptable limit of 0.25 mRem/h. The tally results for the simplified cylindrical model pass all ten of the statistical checks within MCNP6.2, which indicates a high degree of validity of the simulation results. Shielding types 1 & 3, containing either boric acid in water or containing only light water, resulted in similar shielding effectiveness. The dose rates are also in agreement between the P3 zone from the experimental warehouse and the simplified model. The equivalent dose rate limit is attained when approximately 10.5 cm of boric acid in water shielding is used, 39.7 cm of boric acid in concrete shielding is used, or 11.1 cm of light water shielding is used.

Shielding type 3 using light water would be the best option for shielding to protect staff health during extended operation for experiments. It is the least dense and performs similarly to shielding type 1. This shielding type requires less material than boronated concrete, but requires a hermetic container. Though shielding type 1 using boric acid solution in water could be useful due to its relatively low density, it performs similarly to shielding type 3 and is limited by the solubility of boric acid in water. At room temperature, the solubility of boric acid in water is near 5%.[8] Shielding type 2, involving boric acid in concrete, may be more relevant to terrestrial applications. This shielding type allows for conformability to an experiment without concerns for hermiticity. Either light water shielding or boronated concrete will be most effective for radiation shielding while performing physical experiments within a radiation laboratory warehouse.

**Table 1:** Tabular results for the minimum thickness of shielding material required to reduce the radiation dose below acceptable rates for the three neutron sources and three shielding types.

| Neutron Source | Boric Acid in Water Minimum Thickness [cm] | Boric Acid in Concrete Minimum Thickness [cm] | Light Water Minimum Thickness [cm] |
|---|---|---|---|
| 14.1 MeV Neutron Generator, $3x10^8$ n/s | 77.1 | 94.3 | 77.0 |
| 2.45 MeV Neutron Generator, $10^9$ n/s | 27.6 | 78.4 | 27.9 |
| 10-mCi $^{252}Cf$, $4.4x10^7$ n/s | 10.5 | 39.6 | 11.1 |

## IV. CONCLUSIONS

14.1-MeV, 2.45-MeV, and $^{252}Cf$ neutron sources were simulated using nuclear transport codes to estimate an appropriate thickness of boric acid in concrete, boric acid in water, or light water to reduce the equivalent dose below acceptable levels. The simulation results reveal that a type-1 shielding made of boric acid in light water with a thickness of 77.1 cm or more is effective in reducing the radiation dose below 0.25 mRem/hr for an individual located outside of the shielding and only 2 m away from the source for 14.1 MeV neutrons. A shielding thickness of 94.3 cm or more is effective in reducing the radiation dose below this level for type-2 shielding made of boric acid in concrete. A shielding thickness of 77.0 cm or more is effective in reducing the radiation dose below this level for type-3 shielding made of light water. Therefore, this shield design can effectively decrease the radiation dose below the maximum recommended limit. It is likely that type-3 shielding will be employed due to its relatively low density and shielding effectiveness. Shielding type 2 may be advantageous due to its conformability and ease of design.

## V. CONFLICT OF INTEREST

The authors declare that the research was conducted in the absence of any commercial or financial relationships that could be construed as a potential conflict of interest.

## VI. AUTHOR CONTRIBUTIONS

All authors contributed to the conception and design of the research reported here. AG and CL conducted the MCNP6.2 simulations, and all authors calculated the relevant cross-sections as a function of energy using ENDF. CL and AG wrote the first draft of the manuscript. All authors contributed to manuscript revision, read, and approved the submitted version.

## VII. FUNDING

This work was funded by Texas Tech University and in part by the Texas Research Incentive Program.

## VIII. ACKNOWLEDGEMENTS

This work was supported by Texas Tech University Initiative program. The identification of commercial products, contractors, and suppliers within this article is for informational purposes only, and does not imply endorsement by Texas Tech University, their associates, or their collaborators.

## IX. DATA AVAILABILITY STATEMENT

The data, MCNP input files, and methods utilized in this study can be found at https://www.depts.ttu.edu/phas/cees/, and through the Information Technology Division of Texas Tech University.

## X. REFERENCES

1. P. K. Job, Brookhaven National Laboratory, BNL-210945-2019-TECH
2. J. A. Kulesza *et al*., MCNP® Code Version 6.3.0 Theory & User Manual. Los Alamos National Laboratory Tech. Rep. LA-UR-22-30006, Rev. 1. Los Alamos, NM, USA. September 2022.
3. Brown, D., Chadwick, M., Capote, R., Kahler, A., Trkov, A., Herman, M., et al. "ENDF/B-VIII.0: The 8th major release of the nuclear reaction data library with CIELO-project cross sections, new standards and thermal scattering data", Nucl. Data Sheets 148(2018)1.
4. P. Kerr *et al*., Neutron transmission imaging with a portable D-T neutron generator, J. Radiation Detection Technology and Methods (2022) 6(2) 234-243. https://doi.org/10.1007/s41605-022-00315-7
5. M. Fuller *et al*., Long-Lifetime High-Yield Neutron Generators using the DD reaction, International Topical Meeting on Nuclear Research Applications and Utilization of Accelerators (2009)
6. United States nuclear Regulatory Commission (2021). Units of radiation dose https://www.nrc.gov/reading-rm/doc-collections/cfr/part020/part020-1004.html [Accessed March 15, 2023].
7. United States nuclear Regulatory Commission (2021). https://www.nrc.gov/reading-rm/basic-ref/glossary/rem-roentgen-equivalent-man.html [Accessed March 30, 2023].
8. Schubert D (2011) Boron oxides, boric acid, and borates. Kirk Othmer, Ency Chem Technol:1–68. https://doi.org/10.1002/0471238961.0215181519130920.a01.pub3

## XI. SUPPLEMENTARY MATERIALS

Data tables taken from the US NRC used in calculating the equivalent dose.

**Table S1:** Quality factors and fluence per unit dose equivalent for monoenergetic neutrons.

| Neutron energy (MeV) | Quality factor (Q) [a] | Fluence per unit dose equivalent[b] (neutrons $cm^{-2}$ $rem^{-1}$) |
|---:|---:|---:|
| $2.5 \times 10^{-8}$ | 2 | $980 \times 10^{6}$ |
| $1 \times 10^{-7}$ | 2 | $980 \times 10^{6}$ |
| $1 \times 10^{-6}$ | 2 | $810 \times 10^{6}$ |
| $1 \times 10^{-5}$ | 2 | $810 \times 10^{6}$ |
| $1 \times 10^{-4}$ | 2 | $840 \times 10^{6}$ |
| $1 \times 10^{-3}$ | 2 | $980 \times 10^{6}$ |
| $1 \times 10^{-2}$ | 2.5 | $1010 \times 10^{6}$ |
| $1 \times 10^{-1}$ | 7.5 | $170 \times 10^{6}$ |
| $5 \times 10^{-1}$ | 11 | $39 \times 10^{6}$ |
| 1 | 11 | $27 \times 10^{6}$ |
| 2.5 | 9 | $29 \times 10^{6}$ |
| 5 | 8 | $23 \times 10^{6}$ |
| 7 | 7 | $24 \times 10^{6}$ |
| 10 | 6.5 | $24 \times 10^{6}$ |
| 14 | 7.5 | $17 \times 10^{6}$ |
| 20 | 8 | $16 \times 10^{6}$ |
| 40 | 7 | $14 \times 10^{6}$ |
| 60 | 5.5 | $16 \times 10^{6}$ |
| $1 \times 10^{2}$ | 4 | $20 \times 10^{6}$ |
| $2 \times 10^{2}$ | 3.5 | $19 \times 10^{6}$ |
| $3 \times 10^{2}$ | 3.5 | $16 \times 10^{6}$ |
| $4 \times 10^{2}$ | 3.5 | $14 \times 10^{6}$ |